\documentclass[reprint,amsmath,amssymb,aps,prb,superscriptaddress]{revtex4-2}
\usepackage{dcolumn}
\usepackage{bm}
\usepackage[mathlines]{lineno}
\usepackage[utf8]{inputenc}
\usepackage[T1]{fontenc}
\usepackage{booktabs, array, float, tabularx, lipsum, amsmath, amsthm, amssymb, multirow, esvect}
\usepackage{paralist}

\usepackage{newtxtext,newtxmath}
\DeclareSymbolFont{orilargesymbols}{OMX}{cmex}{m}{n}
\DeclareMathSymbol{\orisum}{\mathop}{orilargesymbols}{"50}
\renewcommand\sum\orisum
\renewcommand\geq\geqslant 
\renewcommand\leq\leqslant
\usepackage[cal=zapfc]{mathalpha}  
\usepackage{makecell}
\usepackage{graphicx}
\usepackage{siunitx}
\usepackage[dvipsnames]{xcolor}
\usepackage[version=4]{mhchem}
\usepackage[colorlinks,linkcolor=blue,anchorcolor=blue,citecolor=blue,urlcolor=blue]{hyperref}
\usepackage{orcidlink}

\begin{document}

\title{High-Efficiency Nonrelativistic Charge-Spin Conversion in X-Type Antiferromagnets}
\newcommand{\WUSTa}{\affiliation{State Key Laboratory of Advanced Refractories, \href{https://ror.org/00e4hrk88}{Wuhan University of Science and Technology}, Wuhan 430081, People's Republic of China}}
\newcommand{\WUSTb}{\affiliation{School of Materials Science and Engineering, \href{https://ror.org/00e4hrk88}{Wuhan University of Science and Technology}, Wuhan 430081, People's Republic of China}}
\newcommand{\HBPU}{\affiliation{School of Electrical and Electronic Information Engineering, \href{https://ror.org/01z07eq06}{Hubei Polytechnic University}, Huangshi 435003, People's Republic of China}}
\newcommand{\WHU}{\affiliation{Key Laboratory of Artificial Micro- and Nano-structures of Ministry of Education,\\ School of Physics and Technology, \href{https://ror.org/033vjfk17}{Wuhan University}, Wuhan 430072, People's Republic of China}}

\author{Jiabin Wang}
\thanks{These authors contributed equally to this work}
\WUSTa
\WUSTb
\author{Wancheng Zhang~\orcidlink{0000-0003-1382-9806}}
\thanks{These authors contributed equally to this work}
\HBPU
\author{Zhenhua Zhang~\orcidlink{0000-0003-2148-8239}}
\email[Contact author:]{zzhua@wust.edu.cn}
\WUSTa
\WUSTb
\author{Rui Xiong~\orcidlink{0000-0003-0468-6014}}
\WHU
\author{Yong Liu~\orcidlink{0000-0001-6095-6926}}
\WHU
\author{Zhihong Lu~\orcidlink{0000-0002-6636-0581}}
\email[Contact author:]{zludavid@live.com}
\WUSTa
\WUSTb

\date{\today}

\begin{abstract}
Antiferromagnetic materials with spin splitting have attracted considerable attention for their symmetry-enabled anisotropic spin textures that sustain a zero net magnetization, thereby facilitating efficient spin-current generation. In this work, the highly efficient generation of nonrelativistic spin currents is demonstrated to be facilitated by the distinctive Fermi surface geometry of X-type collinear antiferromagnets. As a prototype conducting X-type antiferromagnet, the Fermi surface of $\beta$-\ce{Fe2PO5} exhibits a distinct $d$-wave altermagnetic characteristic, which compresses into a nearly X-shaped configuration. This results in highly efficient spin currents, achieving a charge-spin conversion efficiency of up to 90\%. Moreover, the spin current polarization is controlled by the orientation of the N\'eel vector. When the N\'eel vector tilts to the out-of-plane direction, an in-plane injected charge current can generate a special spin current component with both spin polarization and propagation along the out-of-plane direction, whose charge-spin conversion efficiency substantially exceeds that of known ferromagnets, altermagnets, noncollinear antiferromagnets, and low-symmetry materials. The highly efficient charge-spin conversion in X-type antiferromagnets provides a novel and highly effective spin source system for the development of low-power spintronic devices.
\end{abstract}

\maketitle

\setlength {\parskip} {0pt}
\section{Introduction}\label{intro}
Spintronics, which exploits both the charge and spin degrees of freedom of electrons, has emerged as a pivotal frontier in the development of next-generation information technologies~\cite{1,2}. Antiferromagnets have recently emerged as a compelling materials platform for high-performance spintronic devices, owing to their vanishing net magnetic moment, absence of stray fields, and capability for ultrafast THz-scale magnetic dynamics~\cite{3,4}. These attributes address key challenges for post-Moore's Law electronics-including miniaturization, high-density integration, low power operation, high-speed read/write functionality, and robust stability-making them ideal candidates for next-generation information technologies. Antiferromagnets, long perceived as functionally limited materials, have undergone a paradigm shift with the recent discovery of time-reversal symmetry ($\cal T$) breaking in specific classes. This phenomenon induces strong momentum-dependent spin polarization in the electronic band structure, unlocking previously inaccessible spintronic functionalities~\cite{5,6}. These functionalities, previously characteristic of ferromagnetic systems, now emerge in antiferromagnetic materials~\cite{7,8,9,10}.

This class of collinear antiferromagnets, termed altermagnets~\cite{11}, is characterized by spin sublattices interconnected through rotational symmetries, including both proper and improper operations, as well as symmorphic and non-symmorphic transformations~\cite{12}. Theoretical predictions indicate that altermagnets exhibit a novel type of spin splitting mediated by long-range magnetic order, rather than by spin-orbit coupling (SOC)~\cite{13,Z2}. This mechanism enables the generation of spin currents~\cite{7} and the tunnel magnetoresistance effect~\cite{10,11,14} even in light-element compounds, significantly broadening the material candidates for antiferromagnetic spintronics. Most identified altermagnetic materials to date, such as $\alpha$-MnTe~\cite{15}, ${\rm MnTe}_2$~\cite{16}, and recently proposed two-dimensional twisted van der Waals assemblies~\cite{17}, are semiconductors or insulators, which limit their applicability in charge-conducting spintronic devices. Metallic altermagnets remain scarce; among them, $d$-wave \ce{RuO2}~\cite{8,13} and $g$-wave CrSb~\cite{18} are prototypical. However, $g$-wave CrSb lacks the symmetry conditions required for generating nonrelativistic spin-conserving currents. Furthermore, the existence of spin splitting in $d$-wave \ce{RuO2}~\cite{19,20} and the antiferromagnetic nature of its ground state~\cite{21,22,23} remain under active debate, underscoring the need for further fundamental and materials-oriented investigations.

A new class of antiferromagnets characterized by a cross-chain structure, termed X-type antiferromagnets, has been proposed recently~\cite{24}. These systems consist of two magnetic sublattices stacked such that their intersecting atomic chains form an X-shaped pattern. When an electric field is applied along a specific chain, for transport in that direction, electrons with spin orientation aligned with the sublattice magnetization become conductive, whereas those with opposite spin orientation remain insulating. As a result, the spin torque induced by the current acts selectively on only one sublattice, enabling efficient writing and reading of information in antiferromagnetic spintronics. In the representative material $\beta$-\ce{Fe2PO5}, X-type stacking gives rise to sublattice-selective spin-polarized transport in one magnetic sublattice, while the other sublattice remains inactive in spin transport. This unique property allows the exploitation of individual magnetic sublattices within the antiferromagnet, offering a promising route toward high-performance antiferromagnetic spintronic devices.

In this study, we propose that the unique Fermi surface geometry of X-type antiferromagnets enables highly efficient generation of nonrelativistic spin currents, with a charge-spin conversion efficiency approaching 100\%. For (001)-oriented $\beta$-\ce{Fe2PO5}, the Fermi surface exhibits a square-shaped structure with alternating spin textures. Upon a \SI{45}{\degree} rotation about the z-axis (i.e., reorientation to the (110) direction), this structure evolves into a compressed, approximately X-shaped configuration characteristic of $d$-wave altermagnetism. This distinctive Fermi surface geometry leads to highly efficient $\cal T$-odd spin currents with a charge-spin conversion efficiency reaching 90\%. To achieve out-of-plane spin currents in the X-type antiferromagnet, a (101)-oriented unit cell is designed where the N\'eel vector is oriented along the [001] axis. This configuration maintains a remarkable charge-spin conversion efficiency of 80\%, substantially surpassing all known material systems, including ferromagnets, altermagnets, noncollinear antiferromagnets, and low-symmetry crystals. The highly efficient charge-spin conversion in X-type collinear antiferromagnets establishes a material-specific pathway for developing advanced spintronic devices with enhanced performance characteristics.

\setlength {\parskip} {0pt}
\section{methods}\label{math}
The density functional theory is used to implement our first-principles computations in the Vienna ab initio Simulation Package (VASP)~\cite{25}. Pseudopotentials describe the ion-electron interaction, with Perdew-Burke-Ernzerhof (PBE) and generalized gradient approximation (GGA) used for the exchange-correlation potentials~\cite{26,27}. The plane-wave basis cutoff energy is 520 eV, and the convergence criteria for energy and residual force in structure optimization are $1\times10^{-5}$ eV and 0.01 eV/\AA, respectively. During structural optimization, the Brillouin zone (BZ) was sampled using a $\Gamma$-centered $5\times5\times2$ Monkhorst-Pack k-point grid. For self-consistent field (SCF) calculations, Monkhorst-Pack grids of dimensions $8\times8\times3$, $5\times5\times3$, and $3\times8\times8$ were employed for the (001)-, (110)-, and (101)-oriented unit cells, respectively, to ensure accurate electronic structure determination. In our first-principles calculations, temperature effects are not taken into account, and all results are obtained at 0 K.

Hamiltonian for the system was built using Wannier interpolation, with maximally localised Wannier functions derived from the Fe-3$d$ ($d_{xy}, d_{xz}, d_{yz}, d_{z^2}, d_{x^2-y^2}$) and O-2$p$ ($p_x, p_y, p_z$) orbitals. Metals' reaction to electric fields aligns with linear response theory. The $\cal{T}$-odd spin conductivity and spin-charge conversion ratio are assessed within this framework using the Kubo formula, assuming a constant scattering rate $\Gamma$. This analysis is conducted using the \textsc{wannier-linear-response} code~\cite{7,29-SST2,30-WLR1}.

Disorder is thought only to provide a constant band broadening effect in the continuous $\Gamma$ approximation, which modifies the ideal periodic system's Green's functions as follows: $G^R(\varepsilon)=1/(\varepsilon-\hat{H}+i0+) \to 1/(\varepsilon-\hat{H}+i \Gamma)$, where $G^R$ stands for the retarded Green's function, $\varepsilon$ for energy, and $\hat{H}$ for the Hamiltonian ~\cite{31-PhysRevB.77.165117}. Under the constant $\Gamma$ approximation, the Kubo formula can be divided into the contribution~\cite{32-T-odd}
\begin{equation}
  \begin{aligned}
    \label{eq:T_odd}
    \sigma_{\alpha\beta}^{\gamma\text{,odd}} = &-\frac{e\hbar}{\pi}\int \frac{d^3\bm{k}}{(2\pi)^3}\sum_{ n,m} \\
    &\times \frac{\Gamma^2 \text{Re}
    (\langle \psi_{n\bm{k}}| \hat{A}|\psi_{m\bm{k}}\rangle  \langle \psi_{m\bm{k}}| \hat{v}_\beta | \psi_{n\bm{k}}\rangle )}{[(E_F-\varepsilon_{n\bm{k}})^2+\Gamma^2]
    [(E_F-\varepsilon_{m\bm{k}})^2+\Gamma^2]}\text{,} 
  \end{aligned}
\end{equation}


\noindent where $\alpha$, $\beta$, and $\gamma$ stand for the spin current, electric field, and spin polarization directions, respectively, and $e$ is the elementary charge. The quantities $|\psi_{n\bm{k}}\rangle$ and $\varepsilon_{n\bm{k}}$ correspond to the spinor Bloch eigenstates and eigenenergies of the Hamiltonian in the presence of spin-orbit coupling. The summation over the band index $n$ encompasses all spin-orbit split bands, thereby implicitly accounting for the spin degree of freedom. The velocity operator is denoted by $\hat{v}_\beta$, and the Fermi energy is denoted by $E_F$. By defining the operator $\hat{A}=\hat{j}_{\alpha}^{\gamma}$, where $\hat{j}_{\alpha}^{\gamma}=\frac{1}{2}\{\hat{s}_{\gamma}$,$\hat{v}_{\alpha}\}$ is the spin current operator and $\hat{s}_{\gamma}=\frac{\hbar}{2}\hat{\sigma}_{\gamma}$ is the spin operator, Eq.~\eqref{eq:T_odd} can characterise the spin conductivity. To compute the charge conductivity, Eq.~\eqref{eq:T_odd} can be further modified by changing the left side of the equation to $\sigma_{\alpha\beta}$ and setting the operator $\hat{A}=-e\hat{v}_{\alpha}$~\cite{Z2,29-SST2}. By comparing predicted conductivity with observed conductivity, one may estimate the constant $\Gamma$, which determines the magnitude of broadening when matrix elements are transformed by time reversal, an antiunitary operator, Eqs. (1) change differently. As an antiunitary operator, time reversal transforms the matrix elements according to $\langle \psi_{n\bm{k}}|\hat{A} | \psi_{m\bm{k}} \rangle$ $\to$ $\langle \psi_{n\bm{k}}|\mathcal{T} \hat{A} \mathcal{T}| \psi_{m\bm{k}}\rangle^{*}$, thereby leading to distinct transformations of Eq.~\eqref{eq:T_odd} under its operation. Because the spin current operator incorporates an additional spin operator that is $\cal{T}$-odd under time reversal, conductivity and spin conductivity undergo different transformations under time reversal.

\begin{figure}
  \centering
  \includegraphics[width=0.5\textwidth]{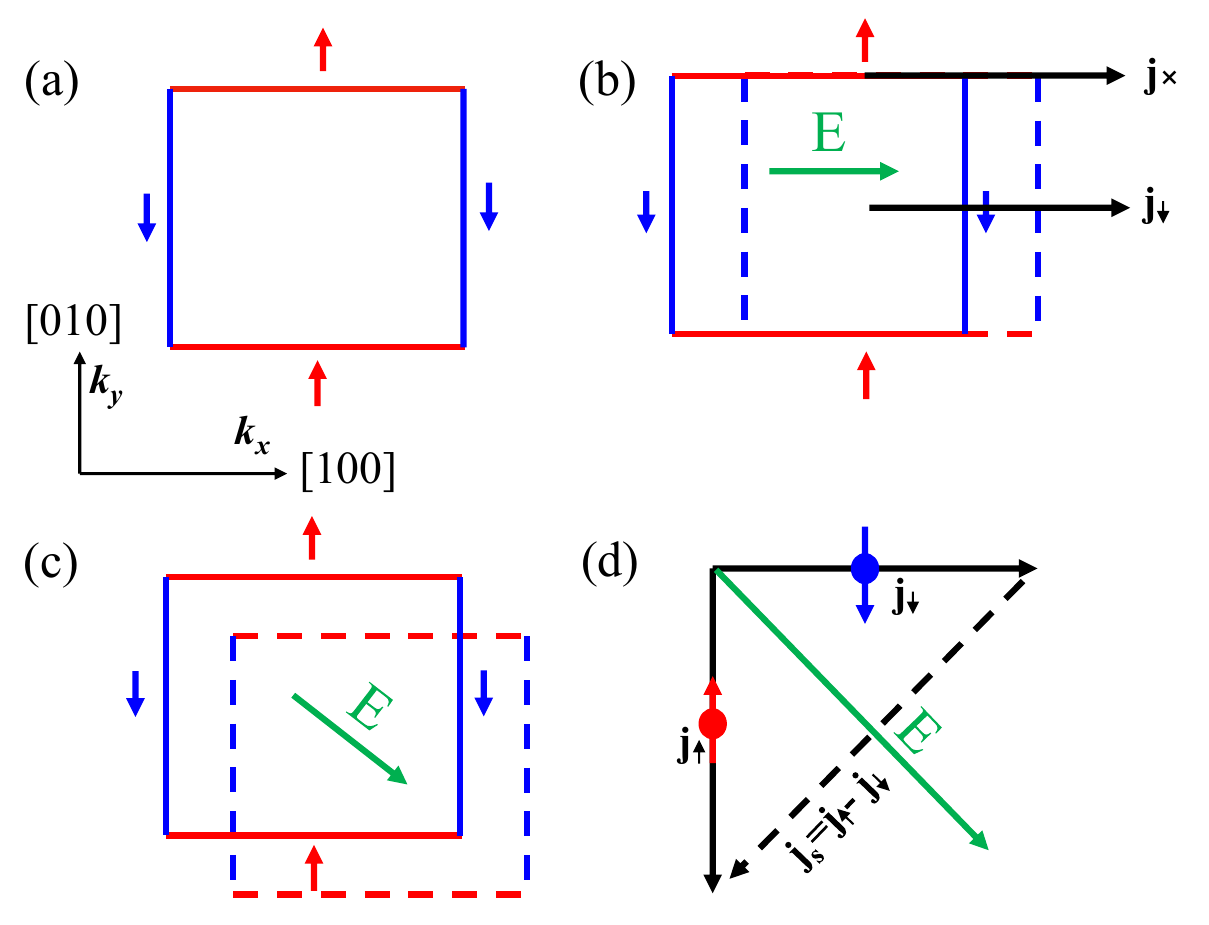}
  \caption{\label{Fig1}Schematic illustration of anisotropic Fermi surface splitting induced by X-type collinear antiferromagnetic order. (a) Fermi surface structure in equilibrium, showing orthogonal spin-up (red) and spin-down (blue) contours. (b) Under an electric field $E\parallel[100]$, a transverse charge current forms for spin-up carriers, while spin-down carriers remain immobile. (c) Evolution of Fermi surface splitting under $E\parallel[110]$. (d) Charge currents for spin-up and spin-down carriers become equal in magnitude but perpendicular in direction under $E\parallel[110]$, generating a transverse spin current.
  }
\end{figure}

\begin{figure*}
  \centering
  \includegraphics[width=0.95\textwidth]{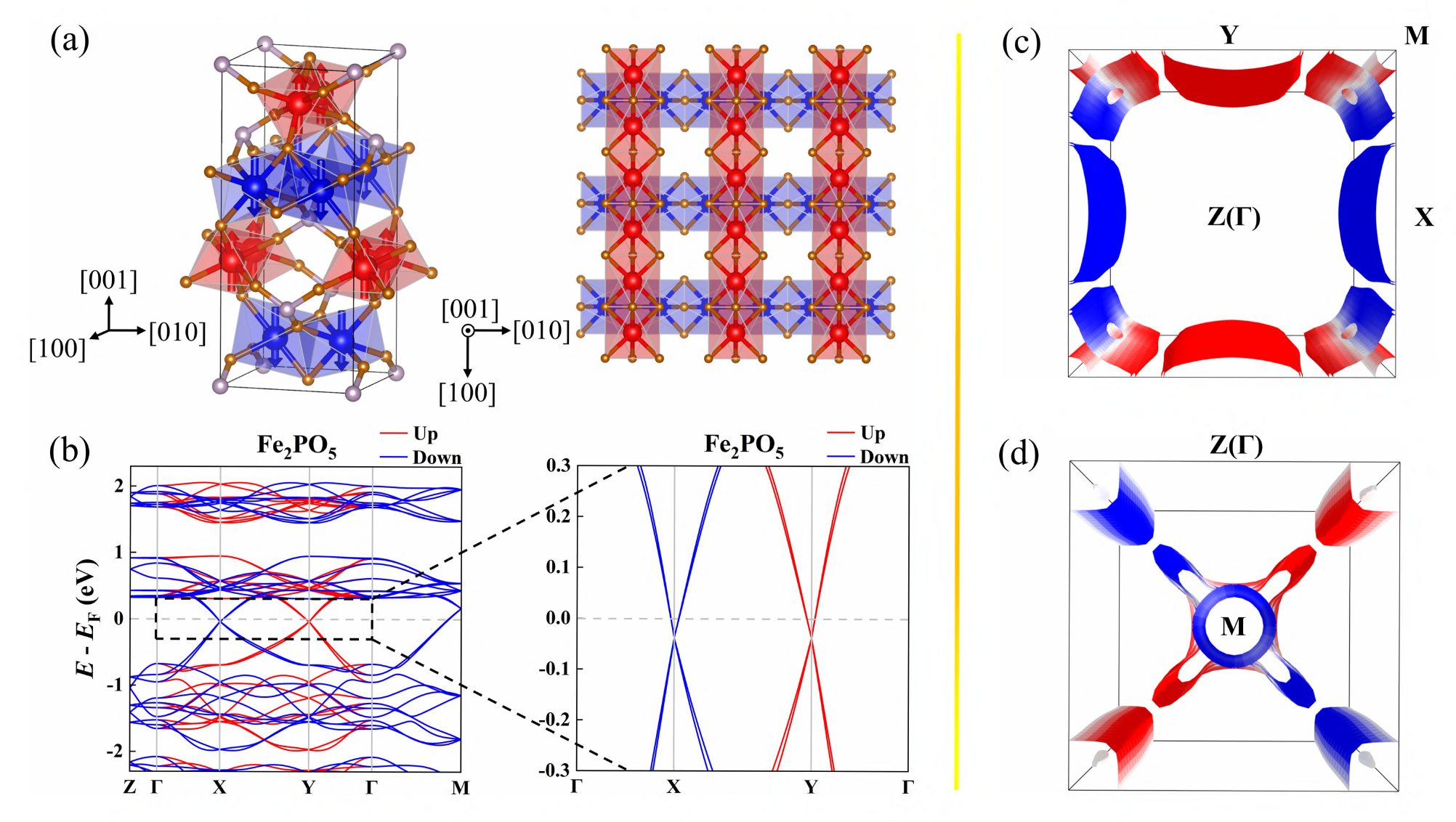}
  \caption{\label{Fig2}Electronic structure of X-type antiferromagnetic $\beta$-\ce{Fe2PO5}. (a) Left: crystal and magnetic structures. Right: Top view of two bottom layers within the unit cell shown on the left, P atoms are omitted for clarity; (b) Electronic band structure with an enlarged view near the $E_{\rm F}$. Spin-resolved Fermi surfaces on the (c) (001)- and (d) (110)-oriented unit cells, with red and blue contours representing spin-up and spin-down channels, respectively.
  }
\end{figure*}

\setlength {\parskip} {0pt}
\section{results and discussion}\label{rnd}
Altermagnetic materials are characterized by nonrelativistic alternating spin splitting in their band structures while maintaining compensated magnetic moments in real space. The low symmetry of $d$-wave altermagnets, such as \ce{RuO2}, \ce{KV2Se2O}~\cite{28}, and \ce{Cr2X2O} (X = Se and Te)~\cite{29} gives rise to anisotropic spin-momentum coupling and spin-dependent conductivity. In X-type collinear antiferromagnetic, the Fermi surface is more distinctive. As shown in Fig.~\hyperref[Fig1]{\ref{Fig1}(a)}, unlike the elliptical Fermi contours of $d$-wave altermagnets, the Fermi surfaces of X-type antiferromagnets such as \ce{Ca(CoO2)2}, $\beta$-\ce{Fe2PO5}, \ce{Co2PO5}, and \ce{VPbO2}~\cite{24} project onto the $xy$-plane as a square geometry, where spin-up and spin-down Fermi surfaces alternate along the square edges (The Fermi surface of \ce{RuO2} is shown in Fig. S1 of the Supplemental Material~\cite{supplement}). When an electric field $E$ is applied along [100] direction, a transverse charge current emerges only for spin-up carriers, while spin-down carriers remain inert, resulting in no net spin current [Fig.~\hyperref[Fig1]{\ref{Fig1}(b)}]. In contrast, under $E\parallel[110]$, the Fermi surfaces shift uniformly, driving a non-spin-polarized charge current along the field direction and a pure spin current in the perpendicular direction [Fig.~\hyperref[Fig1]{\ref{Fig1}(c)} and \hyperref[Fig1]{\ref{Fig1}(d)}]. These characteristics suggest that the X-type antiferromagnet achieves an exceptionally high charge-spin conversion ratio, substantially surpassing that of \ce{RuO2}. In the following, the X-type collinear antiferromagnetic $\beta$-\ce{Fe2PO5} is taken as a representative system to quantitatively validate these predictions. 

\begin{table*}[htbp]
    \caption{\label{tab1}Symmetry analysis of the spin conductivity tensor for $\beta$-\ce{Fe2PO5} with N\'eel vector oriented along the $z$ axis in the presence of SOC, evaluated in the (001)- and (110)-oriented unit cells.}
    \begin{ruledtabular}
        \begin{tabular}{ccccc}
        Orientation & $\mathcal{T}$  & $\sigma^{x}$  & $\sigma^{y}$  & $\sigma^{z}$ \\ \hline & \\[-1.5ex]
        \multirow{5}{*}{(001)-}  &
        $\mathcal{T}$-even  & 
        $\left(\begin{matrix}0&0&0\\0&0&\sigma_{yz}^x=-\sigma_{xz}^y\\0&\sigma_{zy}^x=-\sigma_{zx}^y&0\\\end{matrix}\right)$ & 
        $\left(\begin{matrix}0&0&\sigma_{xz}^y\\0&0&0\\\sigma_{zx}^y&0&0\\\end{matrix}\right)$ & 
        $\left(\begin{matrix}0&\sigma_{xy}^z=-\sigma_{yx}^z&0\\\sigma_{yx}^z&0&0\\0&0&0\\\end{matrix}\right)$ \\ [20pt]
        & 
        $\mathcal{T}$-odd & 
        $\left(\begin{matrix}0&0&\sigma_{xz}^x\\0&0&0\\\sigma_{zx}^x&0&0\\\end{matrix}\right)$ & 
        $\left(\begin{matrix}0&0&0\\0&0&\sigma_{yz}^y=-\sigma_{xz}^x\\0&\sigma_{zy}^y=-\sigma_{zx}^x&0\\\end{matrix}\right)$ & 
        $\left(\begin{matrix}\sigma_{xx}^z&0&0\\0&\sigma_{yy}^z=-\sigma_{xx}^z&0\\0&0&0\\\end{matrix}\right)$  \\ [20pt]
        \multirow{5}{*}{(110)-}  &
        $\mathcal{T}$-even  &
        $\left(\begin{matrix}0&0&0\\0&0&\sigma_{yz}^x=-\sigma_{xz}^y\\0&\sigma_{zy}^x=-\sigma_{zx}^y&0\\\end{matrix}\right)$ & 
        $\left(\begin{matrix}0&0&\sigma_{xz}^y\\0&0&0\\\sigma_{zx}^y&0&0\\\end{matrix}\right)$ & 
        $\left(\begin{matrix}0&\sigma_{xy}^z=-\sigma_{yx}^z&0\\\sigma_{yx}^z&0&0\\0&0&0\\\end{matrix}\right)$ \\ [20pt]
        &
        $\mathcal{T}$-odd  &
        $\left(\begin{matrix}0&0&0\\0&0&\sigma_{yz}^x={\sigma}_{xz}^y\\0&\sigma_{zy}^x={\sigma}_{zx}^y&0\\\end{matrix}\right)$ & 
        $\left(\begin{matrix}0&0&\sigma_{xz}^y\\0&0&0\\\sigma_{zx}^y&0&0\\\end{matrix}\right)$ & 
        $\left(\begin{matrix}0&\sigma_{xy}^z=\sigma_{yx}^z&0\\\sigma_{yx}^z&0&0\\0&0&0\\\end{matrix}\right)$ \\ [15pt]
        \end{tabular}
    \end{ruledtabular}
\end{table*} 

The crystal structure of $\beta$-\ce{Fe2PO5} was obtained from the Materials Project database (mp-18830). As shown in Fig.~\hyperref[Fig2]{\ref{Fig2}(a)}, the crystal and magnetic structures are illustrated alongside a top view of two bottom layers within the unit cell. The figure clearly reveals that the $\beta$-\ce{Fe2PO5} lattice consists of two collinear antiferromagnetic sublattices stacked in a crossed configuration. Within each layer, ferromagnetic chains exhibit parallel alignment, while adjacent layers are coupled antiparallelly, giving rise to the characteristic X-type antiferromagnetic pattern. This unique symmetry leads to unconventional band splitting, as illustrated in Fig.~\hyperref[Fig2]{\ref{Fig2}(b)} (the Brillouin zone with high-symmetry points marked is provided in Supplemental Material Fig. S3~\cite{supplement}). Near the Fermi level, only spin-down bands appear at the high-symmetry X point, while spin-up bands emerge exclusively at the Y point, distinct from the splitting behavior observed in conventional altermagnets such as \ce{RuO2}. The resulting transport properties are governed not only by the antiferromagnetic order but also by symmetry differences between the sublattice-derived Fermi surfaces. To visualize this behavior, spin-resolved Fermi surfaces of $\beta$-\ce{Fe2PO5} are calculated. As shown in Fig.~\hyperref[Fig2]{\ref{Fig2}(c)}, spin-up Fermi arcs extend along the $\Gamma$--X direction, while spin-down arcs propagate along the $\Gamma$--Y direction, consistent with the band structure calculations. However, no intertwined spin-resolved Fermi surfaces are observed in Fig.~\hyperref[Fig2]{\ref{Fig2}(c)}. We constructed three types of unit cells: one aligned with the principal crystallographic axes ([100], [010], [001]), referred to as the (001)-oriented unit cell; another rotated by \SI{45}{\degree} about the [001] axis, with axes along [$1\bar{1}0$], [110], and [001], referred to as the (110)-oriented unit cell; and a third, derived from the (001)-oriented supercell, with axes along [$10\bar{1}$], [010], and [100], referred to as the (101)-oriented unit cell. By applying a rotation matrix to transform the coordinate system from the (001)- to (110)-oriented unit cell, the spin-resolved Fermi surfaces of $\beta$-\ce{Fe2PO5} were recalculated. As shown in Fig.~\hyperref[Fig2]{\ref{Fig2}(d)}, the resulting Fermi surface in the (110)-oriented frame exhibits a compressed X-shaped $d$-wave structure, similar to altermagnetic behavior. This finding explains the emergence of a spin-unpolarized charge current and a transverse spin current under an electric field applied along the [110] direction, as schematically depicted in Fig.~\hyperref[Fig1]{\ref{Fig1}(d)}.

\begin{figure}[htbp]
  \centering
  \includegraphics[width=0.45\textwidth]{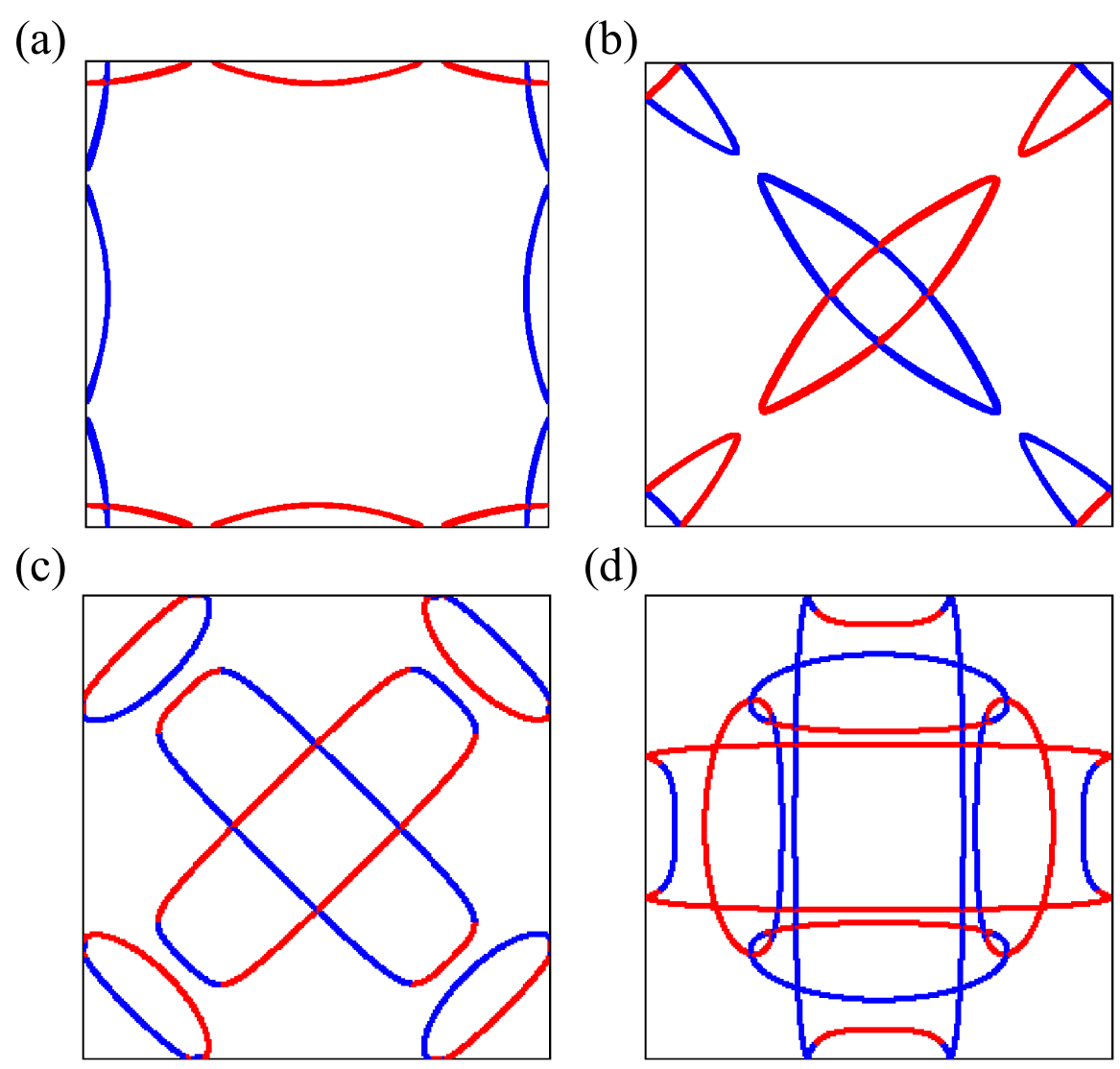}
  \caption{\label{Fig3}Spin-resolved Fermi surface projections. Projections on the $k_z = \pi/2c$ plane for (a) $\beta$-\ce{Fe2PO5} in the (001)-oriented unit cell and (b) $\beta$-\ce{Fe2PO5} in the (110)-oriented unit cell. Projections on the $k_z = 0$ plane for (c) \ce{RuO2} in the (001)-oriented unit cell and (d) \ce{RuO2} in the (110)-oriented unit cell. Red and blue contours represent spin-up and spin-down states, respectively.
  }
\end{figure}

\begin{figure*}[htbp]
  \centering
  \includegraphics[width=0.95\textwidth]{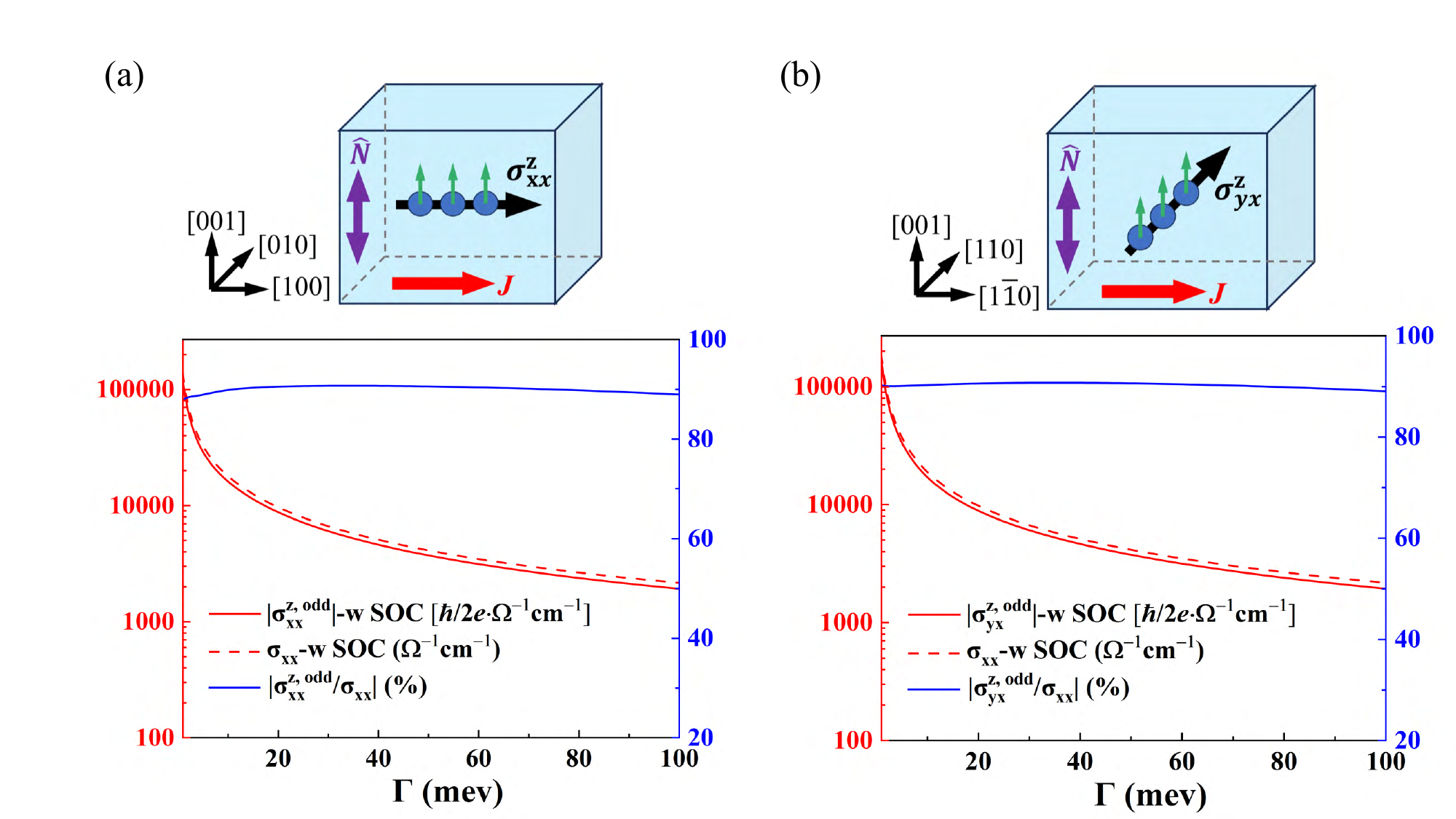}
  \caption{\label{Fig4}Charge-spin conversion response of $\beta$-\ce{Fe2PO5} as a function of scattering rate $\Gamma$. (a) For the (001)-oriented unit cell: $\cal T$-odd spin conductivity $|\sigma_{xx}^{z\rm ,odd}|$ (solid red), charge conductivity $\sigma_{xx}$ (dashed red), and charge-spin conversion efficiency $|\sigma_{xx}^{z\rm ,odd}/\sigma_{xx}|$ (solid blue). (b) For the (110)-oriented unit cell: corresponding $\cal T$-odd spin conductivity $|\sigma_{yx}^{z\rm ,odd}|$, conductivities $\sigma_{xx}$, and conversion efficiency $|\sigma_{yx}^{z\rm ,odd}/\sigma_{xx}|$. Schematic three-dimensional diagrams (top) illustrate the N\'eel vector $\hat{N}$ (purple double arrow), charge current $J$ (red arrow), spin polarization direction (green arrow), and spin current flow (black arrow).
  }
\end{figure*}

To more clearly illustrate the Fermi-surface evolution from the (001)- to the (110)-oriented unit cells in the X-type antiferromagnet $\beta$-\ce{Fe2PO5},  cross-sectional plots on the $k_z = \pi/2c$ plane are calculated and presented in Figs.~\hyperref[Fig3]{\ref{Fig3}(a)} and \hyperref[Fig3]{(b)}. The spin-resolved Fermi contours are consistent with the three-dimensional spin-resolved Fermi surfaces shown in Figs.~\hyperref[Fig2]{\ref{Fig2}(d)} and \hyperref[Fig2]{(e)}. A comparison of the calculated Fermi surfaces of $\beta$-\ce{Fe2PO5} with those of \ce{RuO2}, a prototypical $d$-wave altermagnet, confirms that $\beta$-\ce{Fe2PO5} also exhibits the characteristic features of a $d$-wave altermagnet. Within a broader classification framework, $\beta$-\ce{Fe2PO5} belongs to the Type-I category of the X-type antiferromagnetic family. The three-dimensional spin-resolved Fermi surfaces of \ce{RuO2} for the (001)- and (110)-oriented unit cells are provided in Fig. S1 of the Supplemental Material~\cite{supplement}; the corresponding $k_z = 0$ cross sections are shown in Figs.~\hyperref[Fig3]{\ref{Fig3}(c)} and \hyperref[Fig3]{(d)}. In both $\beta$-\ce{Fe2PO5} and \ce{RuO2}, alternating spin-resolved Fermi surfaces are clearly resolved for both crystallographic orientations. As illustrated in Fig.~\hyperref[Fig3]{\ref{Fig3}(a)}, the Fermi surface of $\beta$-\ce{Fe2PO5} in the (001)-oriented unit cell exhibits a nearly square-shaped alternating pattern. This feature is characteristic of the strongly anisotropic spin-resolved transport behavior inherent to X-type antiferromagnets. Similar Fermi surface morphologies are also observed in Type-II X-type antiferromagnets; however, in these materials, the lifting of band degeneracy near the $\Gamma$ point results in a compensated antiferromagnetic state that is not altermagnetic. Type-III X-type antiferromagnets, which exhibit more complex magnetic structures, generally do not fall within the altermagnetic classification and are not discussed further in this work. $\beta$-\ce{Fe2PO5} was selected as the primary focus of this study due to the strongly anisotropic spin-polarized transport behavior arising from its X-type antiferromagnetic structure. Importantly, this property is not exclusive to Type-I materials but also extends to Type-II X-type antiferromagnets, thereby offering broader physical relevance. The distinctive Fermi surface topology inherent to X-type antiferromagnets holds promise for facilitating efficient spin current generation and charge-spin conversion. Type-III X-type antiferromagnets, which exhibit more complex magnetic structures, generally do not fall within the altermagnetic classification. $\beta$-\ce{Fe2PO5} was selected as the primary focus of this study due to the strongly anisotropic spin-polarized transport behavior arising from its X-type antiferromagnetic structure. Importantly, this property is not exclusive to Type-I materials but extends to Type-II and Type-III X-type antiferromagnets as well, thereby offering broader physical relevance. The distinctive Fermi surface topology inherent to X-type antiferromagnets holds promise for facilitating efficient spin current generation and spin-charge conversion.

To clarify the relationships among the components of the spin conductivity tensor in $\beta$-\ce{Fe2PO5}, a systematic symmetry analysis of the tensor was performed. Table~\ref{tab1} summarizes the third-rank spin conductivity tensor $\sigma_{\alpha\beta}^{\gamma}$ for antiferromagnetic $\beta$-\ce{Fe2PO5} with N\'eel vector aligned along the [001] direction, evaluated in both (001)- and (110)-oriented unit cells. Here, $\alpha$, $\beta$, and $\gamma$ denote the spatial directions of the spin current, electric field, and spin polarization, respectively. The SOC-dependent $\cal T$-even tensor components are characterized by the conventional spin Hall conductivity (CSHC, with $\sigma_{xy}^z=-\sigma_{yx}^z$, $\sigma_{zy}^x=-\sigma_{zx}^y$, and $\sigma_{yz}^x=-\sigma_{xz}^y$). The $\cal T$-even response vanishes entirely in the absence of SOC, as shown in Table SI of the Supplemental Material~\cite{supplement}. This behavior confirms the relativistic origin of conventional spin Hall effect. For the $\cal T$-odd tensor components in the (001)-oriented unit cell, the magnetic spin Hall effect (MSHE) is characterized by $\sigma_{yz}^{y\rm ,odd}=-\sigma_{xz}^{x\rm ,odd}$ and $\sigma_{zy}^{y\rm ,odd}=-\sigma_{zx}^{x\rm ,odd}$. These relations indicate the presence of a ferromagnetic spin Hall effect-like component in X-type antiferromagnetic systems, originating from symmetry breaking induced by exchange interaction~\cite{Z1}. However, since the MSHE strongly depends on the strength of SOC, its magnitude is expected to be small in materials such as $\beta$-\ce{Fe2PO5}, which exhibits relatively weak SOC. Consequently, this type of anomalous spin Hall effect is not further discussed in the present work. Among these tensor components, $\sigma_{xx}^{z\rm ,odd}$ dominates the response while other spin conductivity elements remain negligible. The spin-polarized current generated along the axial direction by the component $\sigma_{xx}^{z\rm ,odd}$ aligns with the Fermi surface analysis presented in Fig.~\hyperref[Fig1]{\ref{Fig1}(b)}. More importantly, we propose the mechanism through which the strong anisotropy in the spin-diagonal conductivities $\sigma_{xx}^{z\rm ,odd}$ and $\sigma_{yy}^{z\rm ,odd}$ gives rise to a large spin Hall conductivity, that is, the generation of a transverse spin current. The transverse spin current will show up by rotating the applied electric field from the [100] to the [110] crystallographic direction. To perform a symmetry analysis of the spin Hall conductivity under an electric field applied along the [110] direction, a unit cell whose principal axes are aligned along the diagonal direction of the (001)-oriented unit cell in Fig.~\hyperref[Fig2]{\ref{Fig2}(a)} is selected, that is the (110)-oriented unit cell. It is found that for the (110)-oriented unit cell, the tensor components satisfy $\sigma_{yz}^{x\rm ,odd}={\sigma}_{xz}^{y\rm ,odd}$, $\sigma_{zy}^{x\rm ,odd}={\sigma}_{zx}^{y\rm ,odd}$, and $\sigma_{xy}^{z\rm ,odd}=\sigma_{yx}^{z\rm ,odd}$. Notably, a nonzero component $\sigma_{xy}^{z\rm ,odd}=\sigma_{yx}^{z\rm ,odd}$ persists even when SOC is neglected.This indicates that applying an electric field along the [110] direction leads to the generation of a transverse pure spin current. This behaviour demonstrates that efficient charge-spin conversion can be achieved in X-type antiferromagnets when charge currents are directed along specific crystallographic axes, even in materials characterised by weak SOC. 

\begin{figure*}[htbp]
  \centering
  \includegraphics[width=0.95\textwidth]{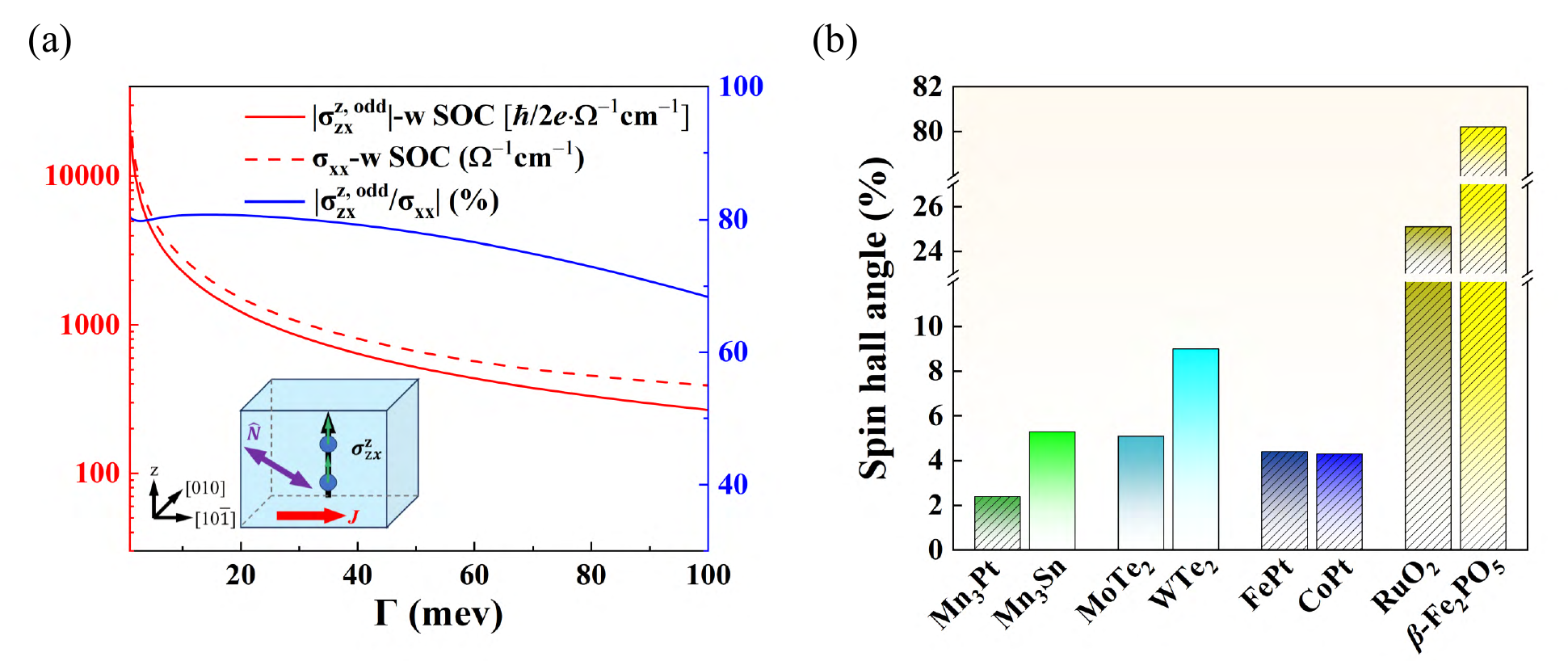}
  \caption{\label{Fig5}(a) The $\beta$-\ce{Fe2PO5} of (101)-oriented unit cell: corresponding $\cal T$-odd spin conductivity $|\sigma_{zx}^{z\rm ,odd}|$, conductivities $\sigma_{xx}$ and conversion efficiency $|\sigma_{zx}^{z\rm ,odd}/\sigma_{xx}|$. Schematic three-dimensional diagrams illustrate the N\'eel vector $\hat{N}$ (purple double arrow), charge current $J$ (red arrow), spin polarization direction (green arrow), and spin current flow (black arrow). (b) charge-spin conversion efficiency for \ce{Mn3Pt}, \ce{Mn3Sn}, \ce{MoTe2}, \ce{WTe2}, FePt, CoPt, \ce{RuO2}, and $\beta$-\ce{Fe2PO5} with out-of-plane polarized spin currents. The bars filled with diagonal lines are obtained from calculations, while those without diagonal lines are obtained from experiments.
  }
\end{figure*}

First-principles calculations of the spin Hall conductivity are typically constrained to the principal axes of the chosen unit cell; therefore, accessing transport along the [110] direction necessitates the use of a (110)-oriented unit cell. Within the framework of the linear-response formula [Eq. (1)]~\cite{supplement}, the spin conductivities $\sigma_{xx}^{z\rm ,odd}$ and $\sigma_{yx}^{z\rm ,odd}$ of $\beta$-\ce{Fe2PO5} are calculated under conditions $\hat{A}=\hat{j}_\alpha^\gamma$ in the (001)- and (110)-oriented unit cells, respectively. The charge conductivity $\sigma_{xx}$ is also evaluated under conditions $\hat{A}=-e\hat{v}_\alpha$ for both orientations~\cite{Z1,Z2,Z3}. For consistent comparison, the magnitudes of both the spin conductivity and charge-spin conversion efficiency are presented. As shown in Fig.~\hyperref[Fig4]{\ref{Fig4}(a)}, in (001)-oriented $\beta$-\ce{Fe2PO5} under an electric field applied along the [100] direction, a spin-polarized current is generated along the same direction. From the ratio of the spin Hall conductivity to the charge conductivity, it is found that when the electric field is applied along the [100] direction, the polarization of the spin-polarized current can reach up to 90\%, which is consistent with the aforementioned band-structure calculations and Fermi-surface analysis in Fig.~\ref{Fig2}. As illustrated in Fig.~\hyperref[Fig4]{\ref{Fig4}(b)}, when the coordinate system is rotated by \SI{45}{\degree} about the [001] N\'eel vector direction, corresponding to the (110)-oriented unit cell, a charge current flowing along [$1\bar{1}0$] generates a spin current along the [110] direction. Notably, the spin conductivity $|\sigma_{xx}^{z\rm ,odd}|$ and the charge-spin conversion efficiency $|\sigma_{xx}^{z\rm ,odd}/\sigma_{xx}|$ in the (001)-oriented unit cell are approximately equal to the spin conductivity $|\sigma_{yx}^{z\rm ,odd}|$ and the charge-spin conversion efficiency $|\sigma_{yx}^{z\rm ,odd}/\sigma_{xx}|$ in the (110)-oriented unit cell. This phenomenon can be understood through the subsequent analysis of the rotation matrix. When contributions from SOC to the spin Hall conductivity matrix elements are neglected, the spin-diagonal conductivity $|\sigma_{xx}^{z\rm ,odd}|$ in the (001)-oriented unit cell is expected, in principle, to be strictly equal to the spin Hall conductivity $|\sigma_{yx}^{z\rm ,odd}|$ in the (110)-oriented unit cell. These results establish that spin transport in the X-type antiferromagnet $\beta$-\ce{Fe2PO5} exhibits pronounced anisotropy. By simply varying the direction of the applied electric field, a fully spin-polarized charge current in $\beta$-\ce{Fe2PO5} can be converted into a highly efficient transverse spin current. The charge-spin conversion efficiency reaches a remarkable 90\%, surpassing that of nearly all known materials to date.


The spin current polarization in X-type antiferromagnets is governed by the N\'eel vector orientation. This enables the generation of out-of-plane spin currents through epitaxial orientation control or direct manipulation of the N\'eel vector, which is essential for practical spintronic applications.
For the (101)-oriented $\beta$-\ce{Fe2PO5} thin film, the [001] axis aligns with the N\'eel vector direction, which tilts out-of-plane as illustrated in the inset of Fig.~\hyperref[Fig5]{\ref{Fig5}(a)}. This configuration enables an in-plane charge current to generate a pure spin current propagating out-of-plane. More importantly, since its spin polarization direction is aligned with the N\'eel vector direction (i.e., the [001] axis), the spin polarization has a component along the $z$-axis in the schematic diagram of Fig.~\hyperref[Fig5]{\ref{Fig5}(a)}. This makes it possible to achieve a pure spin current with both spin polarization and propagation direction out-of-plane (corresponding to $\sigma_{zx}^{z\rm ,odd}$). When injected into an adjacent ferromagnetic layer, this special spin current can achieve perpendicular magnetization switching, which is crucial for the development of high-density magnetic random-access memories (MRAM)~\cite{Z2}. Note that the Fermi surface of $\beta$-\ce{Fe2PO5} is rotated by \SI{45}{\degree} relative to that of \ce{RuO2}, leading to distinct symmetry-constrained spin conductivity tensors in the (101)-oriented unit cell between the two materials. The calculated spin conductivity $\sigma_{zx}^{z\rm ,odd}$, charge conductivity $\sigma_{xx}$, and charge-spin conversion efficiency $|\sigma_{zx}^{z\rm ,odd}/\sigma_{xx}|$ for the (101)-oriented $\beta$-\ce{Fe2PO5} film are presented in Fig.~\hyperref[Fig5]{\ref{Fig5}(a)}, revealing a conversion efficiency as high as 80\%. For comparison, Fig.~\hyperref[Fig5]{\ref{Fig5}(b)} summarizes the charge-spin conversion efficiency of representative material systems, including noncollinear antiferromagnets (\ce{Mn3Pt}, \ce{Mn3Sn})~\cite{30,31}, low-symmetry crystals (\ce{MoTe2}, \ce{WTe2})~\cite{32,33}, ferromagnets (FePt, CoPt)~\cite{34}, and the altermagnet \ce{RuO2} (The charge-spin conversion efficiency of \ce{RuO2} in the (101)-oriented unit cell is presented in Fig. S2 of the Supplemental Material~\cite{supplement}). The charge-spin conversion efficiency of $\beta$-\ce{Fe2PO5} exceeds those of all these established systems. Thus, the out-of-plane polarized spin current with highly efficient generation in X-type antiferromagnets offers a valuable design principle for developing low-power spintronic devices.

In fact, the spin conductivity tensors for the (001)-, (110)-, and (101)-oriented unit cells can be interrelated through the relation $\sigma_{(110/101)}{_{i,j}^{s,k}} =\mathop{\sum}\limits_{l,m,n} D_{il}D_{jm}D_{kn}\sigma_{(001)}{_{l,m}^{s,n}}$~\cite{Z2}. The (110)-oriented unit cell of $\beta$-\ce{Fe2PO5} is obtained by rotating the (001)-oriented frame counterclockwise about the $z$-axis, with the transformation defined by the rotation matrix:
$$
D_1=\begin{pmatrix}
 \cos\varphi & -\sin\varphi & 0 \\
 \sin\varphi &  \cos\varphi & 0\\
 0           &  0           & 1
\end{pmatrix}\text{,}
$$
where for the case of (001)-to-(110)-oriented rotation, $\varphi=\arctan\frac{c}{a}$.

The (101)-oriented unit cell of $\beta$-\ce{Fe2PO5} is obtained by rotating the (001)-oriented frame clockwise about the $y$-axis, with the transformation described by the rotation matrix:
$$
D_2=\begin{pmatrix}
 \cos\varphi  & 0            & \sin\varphi \\
 0            & 1            & 0           \\
 -\sin\varphi & 0            & \cos\varphi
\end{pmatrix}\text{,}
$$
where for the case of (001)-to-(101)-oriented rotation, $\varphi=\arctan\frac{b}{a}$.

Similarly, the charge conductivity tensors for the (110)- and (101)-oriented unit cells are derived using the following transformation: $\sigma_{(110/101)}{_{i,j}} =\mathop{\sum}\limits_{l,m} D_{il}D_{jm}\sigma_{(001)}{_{l,m}}$. Therefore, if the charge conductivity and spin Hall conductivity for the (001)-oriented unit cell are known [Fig.~\hyperref[Fig4]{\ref{Fig4}(a)}], the corresponding values for an arbitrary current direction can be obtained directly through coordinate-system rotation. Accordingly, we can obtain the charge conductivity and spin Hall conductivity for the (110)- and (101)-oriented unit cells via transformations using the matrices $D_1$ and $D_2$ , respectively. It is found that the results obtained from coordinate rotation agree well with those computed from direct first-principles calculations, mutually validating their accuracy (as detailed in Table SII of the Supplemental Material)~\cite{supplement}. In this work, the actual coordinate axis $x$ in the third-rank spin Hall conductivity tensor calculation is oriented opposite to the $x$-axis defined in the rotation-matrix transformation. Consequently, tensor components containing an odd number of $x$ indices undergo a sign reversal under this coordinate transformation.

Notably, such $\cal T$-odd nonrelativistic spin currents are not confined to X-type antiferromagnetic systems. In low-symmetry ferromagnetic or antiferromagnetic systems exhibiting spin splitting, electron transport along specific crystallographic directions can generate spin-polarized currents. When the spin conductivity tensor of such polarized currents exhibits crystallographic anisotropy in magnetic systems (including altermagnets), tensor rotation operations demonstrate that transverse spin currents can be generated along non-principal crystalline axes. The charge-spin conversion efficiency of such transverse spin currents correlates strongly with the anisotropy strength in polarized current transport within the material system. Moreover, the spin polarization direction of these magnetically derived spin currents is governed by the system's order parameter, enabling the generation of out-of-plane polarized spin currents. This mechanism provides a versatile, highly controllable spin-source platform for field-free switching of perpendicularly magnetised MRAM devices. During the submission of this work, a recent study reported a similar Fermi surface structure in an altermagnetic material, resulting in a high charge-spin conversion efficiency (78\%). This efficiency is further enhanced by electron doping~\cite{43-bf1n-sxdl}. Our work indicates that, in a broader range of antiferromagnetic materials, highly efficient charge-spin conversion can be generated through spin-transport anisotropy, as exemplified by the X-type antiferromagnet investigated here, which exhibits a conversion efficiency of up to 90\%. This finding may provide more effective guidance for the practical fabrication of materials.

\setlength {\parskip} {0pt}
\section{conclusions}
This work introduces X-type collinear antiferromagnets distinctive Fermi surface structure that enables the generation of highly efficient $\cal T$-odd spin currents along specific crystallographic directions, outperforming altermagnets in spin-charge conversion, with the spin current polarization controlled by the N\'eel vector orientation. In the prototypical X-type system $\beta$-\ce{Fe2PO5}, the Fermi surface adopts a square geometry with alternating spin textures. When oriented along the [110] direction, the Fermi surface displays a $d$-wave altermagnetic character and compresses into an approximately X-shaped configuration. This unique Fermi surface facilitates highly efficient $\cal T$-odd spin currents, achieving a charge-spin conversion efficiency of 90\%. Moreover, in the (101)-oriented unit cell with the N\'eel vector aligned along the [001] axis, the system generates out-of-plane spin currents while maintaining a conversion efficiency of 80\%, significantly surpassing all known material systems. These results establish X-type collinear antiferromagnets as a promising materials platform for the development of low-power spintronic devices with exceptional spin-charge conversion performance.

\setlength {\parskip} {0pt}
\section*{Acknowledgments}
The authors would like to acknowledge the financial support from National Key Research and Development Program of China (Grant No. 2022YFA1602701), and National Natural Science Foundation of China (Grants No. 12574131, No. 12327806, and No. 12227806). Numerical calculation is supported by High-Performance Computing Center of Wuhan University of Science and Technology. 
Jiabin Wang and Wancheng Zhang contributed equally to this work.

\setlength {\parskip} {0pt}
\section*{Data Availability}
The data that support the findings of this article are openly available~\cite{data}.


%

\end{document}